# Electron Correlation by Exchange Mapping in Electronic Structure Calculations

Jerry L. Whitten

Department of Chemistry

North Carolina State University

Raleigh, NC 27695 USA

email: whitten@ncsu.edu

**Abstract**

A method for increasing the accuracy of configuration interaction (CI) calculations of molecules and other electronic systems is proposed. The energy defect of a given calculation is associated with the electron pair origin of configurations not yet generated and this defect is mapped onto the exchange interaction for the corresponding pair of spatial molecular orbitals. The orbitals can be of opposite spin and thus the contribution includes the self-energy and differs from fermion exchange due to antisymmetry. A single parameter, $\gamma$, multiplying the exchange integral, is determined from the exact thermodynamic energy of a few reference molecules. The value of $\gamma$ depends on the basis and level of configuration interaction but is the same for all molecules. Calculated energies are compared with experimental thermodynamic data for a set of forty mainly organic molecules representing a wide range of bonding environments. Results are reported for two types of multi-reference CI calculations: 1) a triple-zeta basis plus d-type functions for C, N, O and F and an s, p basis for H, and 2) a severely truncated virtual space in which higher spherical harmonic basis functions are removed. The error of the initial CI calculations is large, however, including the exchange-based contribution brings calculated CI energies into much closer agreement with exact values.

I. Introduction

The accuracy of electronic structure calculations by configuration interaction (CI) depends on the completeness of the single particle basis and the completeness of the set of configurations used in the expansion of the wavefunction. As systems increase in size it becomes increasingly difficult to obtain energies close to the exact values. There is a vast literature on CI approaches ranging from perturbation methods that generate configurations and evaluate energies efficiently to methods for partitioning large systems into localized electronic subspaces or ways to balance errors in systems that are being compared.[1-39] Relatively few configurations are required to dissociate molecules correctly or to create proper spin states, but dynamical correlation effects, particularly those associated with angular correlation, require higher spherical harmonic basis functions and this leads to a rapid increase in number of interacting configurations. Finding more efficient ways to treat large systems by coupled-cluster[22-24] and multi-reference methods[9-11,25-27] and methods that use non-orthogonal molecular orbitals[28-29] are continuing research areas in the quest to find accurate descriptions of ground and excited states of molecules and materials.

The present work begins by assuming a CI calculation has been carried out on an electronic system. The calculation is assumed to be sufficiently accurate to describe spin states and capture important static correlation contributions but may be deficient in its single-particle basis or completeness of the CI. The objective of the present work is to explore a new method to account for errors due to the missing configurations.

In the following sections, ideas underlaying the proposed exchange mapping method are discussed and then tested by multi-reference CI calculations on a data set of forty mainly organic molecules containing C, N, O, F and H in a variety of bonding environments. Calculated energies are compared with experimental thermodynamic values. Results are reported for two types of multi-reference CI calculation: 1) calculations that utilize the full occupied and virtual space from an SCF treatment with a triple-zeta basis plus d-type functions for C,N,O and F and an s, p basis for H, and 2) calculations in which the virtual space is severely truncated by removing higher spherical harmonic basis functions necessary for angular correlation.

The present results are also compared with a density method, described in a previous paper by the author (JCP 2020)[5], where the objective was to correct lower-level CI calculations to match more closely the energies of higher-level calculations or the exact energy.

II. Exchange mapping

We begin with a single determinant of spin orbitals $\chi_i$

$$\Phi = (N!)^{-1/2} \det(\chi_{1\,(1)} \chi_{2\,(2)} \chi_{3\,(3)} \cdots \chi_{i\,(i)} \cdots \chi_{j\,(j)} \cdots \chi_{N\,(N)})$$

and the total energy expressed as one-and two- electron contributions

$$E = \sum_i <\chi_i \mid h \mid \chi_i> + \sum_{i<j} <\chi_{i\,(1)}\chi_{j\,(2)} \mid r_{12}^{-1} \mid \chi_{i\,(1)}\chi_{j\,(2)}> - <\chi_{i\,(1)}\chi_{j\,(2)} \mid r_{12}^{-1} \mid \chi_{j\,(1)}\chi_{i\,(2)}>$$

A multi-determinant or CI wavefunction is of the form

$$\Psi = \sum_p c_p (N!)^{-1/2} \det(\chi_1^p{}_{(1)} \chi_2^p{}_{(2)} \chi_3^p{}_{(3)} \cdots \chi_i^p{}_{(i)} \cdots \chi_j^p{}_{(j)} \cdots \chi_N^p{}_{(N)}) = \sum_p c_p \Phi_p$$

where $\chi_i^p$ are chosen from the set of all possible spin orbitals $\{\chi_i\}$. Its energy can be expressed in terms of the same one-and two-electron integrals for single-determinant, $<\Phi_p \mid H \mid \Phi_p>$, plus the additional integrals $<\chi_i \mid h \mid \chi_j>$ and $<\chi_{i\,(1)}\chi_{j\,(2)} \mid r_{12}^{-1} \mid \chi_{k\,(1)}\chi_{l\,(2)}>$ required for matrix elements $<\Phi_p \mid H \mid \Phi_q>\ p \neq q$. The accuracy of the wavefunction is improved by generating additional determinants that interact with the initial expansion.

We now depart from commonly used methods for generating wavefunctions of increasing accuracy by attempting to map the missing electron correlation for a given level of configuration interaction onto exchange-type interactions. Specifically, for excitations that would derive from an $i, j$ pair of spin orbitals in any determinant of the current expansion, the missing correlation energy is set equal to $-\gamma(\chi_{i\,(1)}\chi_{j\,(2)} \mid r_{12}^{-1} \mid \chi_{j\,(1)}\chi_{i\,(2)})$ where the parenthesis notation means that $\alpha, \beta$ spin factors are ignored and only the spatial components of spin orbitals $\chi_i$ and $\chi_i$ occur in the integral. The objective is to obtain the exact energy of the molecule or system by adding the exchange term to the energy expression. Since the exchange integral is positive definite, this obviously can be done for any molecule by appropriate choice of $\gamma$. Since all pairs are included regardless of spin, the contribution can be interpreted as a modification of the Coulomb interaction for each pair of orbitals to give

$$(\chi_{i\,(1)}\chi_{j\,(2)} \mid r_{12}^{-1} \mid \chi_{i\,(1)}\chi_{j\,(2)}>) - \gamma(\chi_{i\,(1)}\chi_{j\,(2)} \mid r_{12}^{-1} \mid \chi_{j\,(1)}\chi_{i\,(2)}) \qquad \gamma \geq 0$$

thereby reducing electron repulsion in the overlap or exchange region. It is not obvious that $\gamma$ would be the same for different molecules. However, we find that rather accurate total energies can be obtained for a wide variety of molecules using one numerical value of $\gamma$. This is true even when the error in the calculated CI energy prior to the exchange contribution is quite large. As the accuracy of the CI improves through larger expansions, one expects $\gamma$ to decrease. It should be noted that the exchange term only shifts the energy of contributing determinants and matrix elements between determinants are unchanged.

By limiting consideration to $i, j$ pairs, we implicitly assume that the CI treatment includes the necessary single excitations from the starting wavefunction. All determinants in the expansion, generated by single or multiple excitations, are subject to the $i,j$ pair energy shift. We emphasize that the exchange function being introduced is not due to fermion exchange but instead is a construction that has the same functional form. In the following, we refer to this as corr-exchange to distinguish it from exchange due to antisymmetry.

## III. Computational details

All calculations are carried out for the full electrostatic Hamiltonian of the system

$$H = \sum_i^N [-\tfrac{1}{2}\nabla_i^2 + \sum_k^Q -\frac{Z_k}{r_{ik}}] + \sum_{i<j}^N r_{ij}^{-1}$$

Calculations on all molecules containing C, N, O and F assume an invariant 1s core; valence orbitals are orthogonal to the core 1s orbitals, but no further mixing of the 1s and valence basis functions is allowed. The 1s core is also invariant in the CI. No core potentials are used

The multi-reference CI procedure used in the present work is described elsewhere, and important details are summarized in the Appendix. SCF orbitals are used for the initial occupied space, and virtual orbitals are determined by a positive ion transformation to improve convergence. Single and double excitations from a single determinant SCF wavefunction, $\Phi_r$, limited by interaction thresholds, creates a small CI expansion, $\Psi'_r$, for state $r$

$$\Psi'_r = \Phi_r + \sum_{ijkl} \lambda_{ijkl} \Gamma_{ij \to kl} \Phi_r = \sum_m c_m \Phi_m$$

The description is then refined by generating a larger CI expansion, $\Psi_r$ by single and double excitations from all important members of $\Psi'_r$ to obtain

$$\Psi_r = \Psi'_r + \sum_m \left[ \sum_{ik} \lambda_{ikm} \Gamma_{i \to k} \Phi_m + \sum_{ijkl} \lambda_{iklm} \Gamma_{ij \to kl} \Phi_m \right]$$

where $\Phi_m$ is a member of $\Psi'_r$. For the molecules investigated, typically, the small expansion $\Psi'_r$ contains 200-400 determinants depending on the molecule, and approximately $10^5$-$10^6$ determinants occur in the final CI expansion, corresponding to single through quadruple excitations from the initial representation of the state $\Phi_r$.

In the present work, the corr-exchange contributions are introduced at the CI stage of the calculation; thus, the SCF and virtual orbital transformation are not affected. As noted above, replacing the Coulomb interaction by

$$(\chi_i(1) \chi_j(2) | r_{12}^{-1} | \chi_i(1) \chi_j(2) >) - \gamma(\chi_i(1) \chi_j(2) | r_{12}^{-1} | \chi_j(1) \chi_i(2))$$

only shifts the energy of $<\Phi_p | H | \Phi_p>$; matrix elements between determinants are not affected.

## IV. Results

Before considering further details, it is instructive to examine how the error of a single-determinant SCF wavefunction, $E_{exact} - E_{SCF}$, scales relative to the total corr-exchange energy, $E_x$, for the same wavefunction

$$E_x = \sum_{i<j} (\chi_i(1)\chi_j(2) | r_{12}^{-1} | \chi_j(1)\chi_i(2))$$

A set of 40 molecules, ordered by size of $E_x$, is considered and Fig. 1 shows a plot of $E_{exact} - E_{SCF}$ vs $E_x$. On this large scale, the relationship is approximately linear and would correspond to $\gamma = 0.036$. Thus, scaling of the correction with size is encouraging.

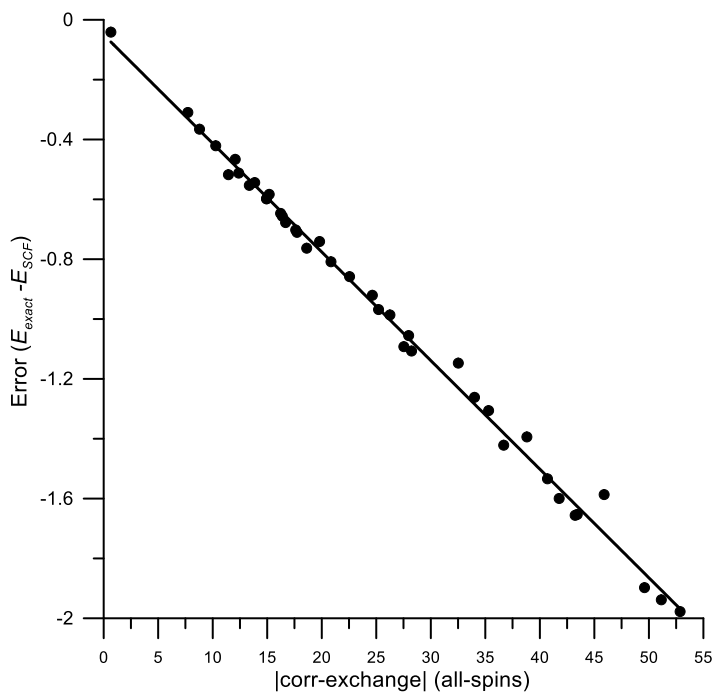

**Fig. 1.** Error vs. corr-exchange. The plot shows $E_{exact} - E_{SCF}$ vs $E_x = \sum_{i<j} (\chi_i(1)\chi_j(2) | r_{12}^{-1} | \chi_j(1)\chi_i(2))$ (all spins) is approximately linear over a wide error range with $slope = -0.036 = \dfrac{\partial(E_{exact} - E_{SCF})}{\partial E_x}$

We now consider details of the calculations. Basis sets are described in Appendix. Hartree-Fock quality expansions of 1s, 2s and 2p orbitals for C, N, O, and F are employed plus additional $s', s'', p', p'', d$ functions to allow polarization and angular correlation contributions (triple-zeta+d basis) ; the basis for H is $1s, s', p'$ (Hsp basis). All calculations begin with a SCF calculation on the molecule of interest using the full basis to provide flexibility and account for polarization contributions. Two different CI calculations are then carried out: 1) with the full basis where virtual orbitals are transformed by a positive ion iteration. 2) with a severely truncated virtual space in which higher spherical harmonic basis functions and the additional polarization function are removed, i.e., removal of $s'', p'', d$ for C, N,O.F and H $s', p'$. The virtual orbitals are again determined by a positive ion transformation. The number of molecular orbitals is greatly reduced in the latter treatment; for example, the number of virtual molecular orbitals is only 12 for $CH_4$ (reduced from 33) and 45 for benzene (reduced from 117). The resulting very small CI expansions provide a test of the effectiveness of increasing the corr-exchange parameter $\gamma$ to account for the larger number of missing correlation excitations compared to the full basis treatment. Truncation of the virtual space can be accomplished either by localization or as in the present work by carrying out a SCF calculation with the unwanted functions removed. The virtual molecular orbitals are then orthogonalized to the occupied SCF orbitals from the full basis calculation and to other virtual orbitals, and then undergo a positive ion transformation. Proceeding in this way avoids confusing the polarization and correlation roles of the omitted functions since the full basis and polarization effects are included in the single-determinant SCF calculation.

The value of the corr-exchange parameter, $\gamma$, is determined by minimizing the percentage error in the atomization energy for a few small molecules containing C, N, O, F with single, double or triple bonds. Since exact energies are used for the atoms, the error in atomization energy of a molecule is the same as the error in the calculated total energy. In Table 1, energy errors are tabulated for CI calculations on the reference molecules with and without the corr-exchange contribution; the calculations will be refined later to minimize the percentage error in atomization. A few larger molecules are included in the table to illustrate typical errors that occur when the same value of $\gamma$ is used for all molecules.

**Table 1.** CI calculations on reference molecules plus results for a few larger molecules to illustrate typical errors that occur when the same value of γ is used for all molecules. Energies are tabulated for uncorrected CI calculations and for calculations including the corr-exchange contribution. Results are reported for two different CI calculations: one, using the full virtual space and a second, smaller CI, using a truncated virtual space. Exact refers to the experimental thermodynamic value.[a]

|  | | virtual space (all s, p, d Hp) | | | | truncated virtuals (no p', d, Hp) | | | |
|---|---|---|---|---|---|---|---|---|---|
|  | exact | CI no corr γ=0 | error | CI γ=0.0176 | error |  |  | γ=0.0254 | error |
| *reference* | | | | | | | | | |
| $H_2$ | -1.17447 | -1.16995 | 0.0045 | -1.18142 | -0.0069 | -1.15668 | 0.0178 | -1.17335 | 0.0011 |
| $CH_4$ | -40.52437 | -40.41127 | 0.1131 | -40.54489 | -0.0205 | -40.32895 | 0.1954 | -40.52347 | 0.0009 |
| $C_2$ | -75.94423 | -75.75317 | 0.1911 | -75.94235 | 0.0019 | -75.67594 | 0.2683 | -75.95138 | -0.0072 |
| $C_2H_2$ | -77.35683 | -77.14305 | 0.2138 | -77.35820 | -0.0014 | -77.04287 | 0.3140 | -77.35592 | 0.0009 |
| $C_2H_4$ | -78.60833 | -78.38975 | 0.2186 | -78.63013 | -0.0218 | -78.25666 | 0.3517 | -78.60666 | 0.0017 |
| $C_2H_6$ | -79.84529 | -79.61900 | 0.2263 | -79.88109 | -0.0358 | -79.46997 | 0.3753 | -79.85320 | -0.0079 |
| $NH_3$ | -56.58549 | -56.43975 | 0.1457 | -56.59244 | -0.0070 | -56.34351 | 0.2420 | -56.56581 | 0.0197 |
| $N_2$ | -109.58714 | -109.32370 | 0.2634 | -109.58386 | 0.0033 | -109.20466 | 0.3825 | -109.58283 | 0.0043 |
| HNNH | -110.6901 | -110.41850 | 0.2716 | -110.70020 | -0.0101 | -110.26931 | 0.4208 | -110.67988 | 0.0102 |
| $N_2H_4$ | -111.9204 | -111.62790 | 0.2925 | -111.93255 | -0.0122 | -111.45330 | 0.4671 | -111.89838 | 0.0220 |
| HOOH | -151.647 | -151.27636 | 0.3706 | -151.63872 | 0.0082 | -151.08913 | 0.5578 | -151.61673 | 0.0302 |
| $H_2O$ | -76.47989 | -76.29405 | 0.1858 | -76.47275 | 0.0071 | -76.19105 | 0.2888 | -76.45095 | 0.0289 |
| $O_2$ | -150.4112 | -150.06074 | 0.3504 | -150.38546 | 0.0257 | -149.90118 | 0.5100 | -150.37299 | 0.0382 |
| HF | -100.5319 | -100.30008 | 0.2318 | -100.51670 | 0.0152 | -100.20155 | 0.3304 | -100.51557 | 0.0163 |
| F2 | -199.675 | -199.21716 | 0.4579 | -199.64732 | 0.0277 | -199.04253 | 0.6325 | -199.66629 | 0.0087 |
| *other molecules* | | | | | | | | | |
| $H_2CO$ | -114.56070 | -114.27178 | 0.2889 | -114.55669 | 0.0040 | -114.12927 | 0.4314 | -114.54366 | 0.0170 |
| $C_6H_6$ | -232.31284 | -231.62659 | 0.6862 | -232.33417 | -0.0213 | -231.31046 | 1.0024 | -232.33896 | -0.0261 |
| $NH_2CH_2COOH$ | -284.59469 | -283.83743 | 0.7573 | -284.59232 | 0.0024 | -283.45912 | 1.1356 | -284.56033 | 0.0344 |
| $NC_4H_4N$ | -264.4243 | -263.66374 | 0.7605 | -264.41541 | 0.0088 | -263.32510 | 1.0991 | -264.41813 | 0.0061 |
| $C_6H_5F$ | -331.6505 | -330.73273 | 0.9178 | -331.65332 | -0.0028 | -330.31036 | 1.3401 | -331.64510 | 0.0054 |
| $C_2F_2H_2$ | -277.2519 | -276.56424 | 0.6877 | -277.23997 | 0.0120 | -276.27748 | 0.9744 | -277.25935 | -0.0074 |

[a] Ref. 2.

From the results in Table 1, we note the following:

1. The initial total energy error in both CI treatments (compared to the exact energy) is quite large for all molecules except hydrogen.
2. Including the corr-exchange contribution with $\gamma = 0.0176$ for the full-basis treatment greatly reduces the error.
3. Increasing the corr-exchange to $\gamma = 0.0254$ for the truncated virtual space CI provides an excellent account of the additional correlation missing in the small CI treatment.
4. The same $\gamma$ values work quite well for the larger molecules.

We now consider refinement. As noted earlier, the 1s core orbitals of C, N, O and F are properly orthogonalized, but invariant in the SCF and CI calculations. This means it can be useful to include atomic corrections to the calculated total energy to account for differences in core correlation energies and other constraints. In order to add a strictly atomic correction, we introduce a new constant, $E_{atomic} = \sum_{k}^{nuclei} \varepsilon_k n_k$, where $\varepsilon_k$ is a constant for nucleus $k$ and $n_k$ is the number of nuclei of type $k$. The constants $\varepsilon_k$ are determined after calculations on a molecule are completed to minimize the percentage error in atomization energy. A small energy shift occurs for H and larger shifts for atoms that have 1s core electrons. Since the energy shifts depend only on the number and type of atoms, these constants do not contribute to the energy change for any chemical reaction that conserves atoms, i.e., the energy change would be the same as calculated by CI with only the corr-exchange contribution. Results, including $\varepsilon_k$, are given in Table 2. Energy errors are found to be slightly reduced compared to the values in Table 1 without the atomic shifts. The atomic constants improve the atomization energy for dissociation to exact atoms. For the full basis calculation, the hydrogen constant in Table 2 is positive, +0.1 eV reflecting the fact that the basis already provides an accurate description of the correlation energy. For the truncated CI calculation, the hydrogen constant is negative, -0.05eV due to the smaller basis.

**Table 2.**  Atomization energies. Results are reported for reference molecules and for other systems using the same value of γ. The CI energies include the corr-exchange contribution and are reported with and without atomic energy shifts (see text). The atomic constants are tabulated in the table. Results are reported for two different CI calculations: one, using the full virtual space and a second, smaller CI using the truncated virtual space. Exact refers to the experimental thermodynamic energy.[a]

|  | Exact [a] |  |  | CI virtual space (all s, p, d Hp) γ=0.0176 |  |  |  | CI truncated virtuals (no p', d, Hp) γ=0.0254 |  |  |  |
|---|---|---|---|---|---|---|---|---|---|---|---|
|  | total E | sum of atomic E | atomization energy | E no shift | E incl shift | error | atomiz error (%) | E no shift | E incl shift | error | atomiz error (%) |
| *reference* |  |  |  |  |  |  |  |  |  |  |  |
| $H_2$ | -1.17447 | -1.00000 | 0.17447 | -1.18142 | -1.17282 | -0.00165 | -0.95 | -1.17335 | -1.17735 | 0.00288 | 1.65 |
| $CH_4$ | -40.52437 | -39.85577 | 0.66860 | -40.54489 | -40.52819 | 0.00382 | 0.57 | -40.52347 | -40.52687 | 0.00250 | 0.37 |
| $C_2$ | -75.94423 | -75.71154 | 0.23269 | -75.94235 | -75.94335 | -0.00088 | -0.38 | -75.95138 | -75.94219 | -0.00204 | -0.88 |
| $C_2H_2$ | -77.35683 | -76.71154 | 0.64529 | -77.35820 | -77.35060 | -0.00623 | -0.97 | -77.35592 | -77.35073 | -0.00610 | -0.95 |
| $C_2H_4$ | -78.60833 | -77.71154 | 0.89679 | -78.63013 | -78.61393 | 0.00560 | 0.62 | -78.60666 | -78.60546 | -0.00287 | -0.32 |
| $C_2H_6$ | -79.84529 | -78.71154 | 1.13375 | -79.88109 | -79.85629 | 0.01100 | 0.97 | -79.85320 | -79.85600 | 0.01071 | 0.95 |
| $NH_3$ | -56.58549 | -56.11160 | 0.47389 | -56.59244 | -56.58100 | -0.00449 | -0.95 | -56.56581 | -56.57716 | -0.00833 | -1.76 |
| $N_2$ | -109.58714 | -109.22320 | 0.36394 | -109.58386 | -109.58678 | -0.00036 | -0.10 | -109.58283 | -109.59353 | 0.00639 | 1.76 |
| HNNH | -110.69009 | -110.22320 | 0.46689 | -110.70020 | -110.69452 | 0.00443 | 0.95 | -110.67988 | -110.69458 | 0.00449 | 0.96 |
| $N_2H_4$ | -111.92037 | -111.22320 | 0.69717 | -111.93255 | -111.91827 | -0.00210 | -0.30 | -111.89838 | -111.91708 | -0.00329 | -0.47 |
| HOOH | -151.64696 | -151.21960 | 0.42736 | -151.63872 | -151.65309 | 0.00613 | 1.43 | -151.61673 | -151.65562 | 0.00866 | 2.03 |
| $H_2O$ | -76.47989 | -76.10980 | 0.37009 | -76.47275 | -76.47564 | -0.00425 | -1.15 | -76.45095 | -76.47239 | -0.00750 | -2.03 |
| $O_2$ | -150.41118 | -150.21960 | 0.19158 | -150.38546 | -150.40843 | -0.00275 | -1.43 | -150.37299 | -150.40788 | -0.00330 | -1.72 |
| HF | -100.53190 | -100.30707 | 0.22483 | -100.51670 | -100.52624 | -0.00566 | -2.52 | -100.51557 | -100.52193 | -0.00997 | -4.43 |
| F2 | -199.67501 | -199.61414 | 0.06087 | -199.64732 | -199.67501 | 0.00000 | 0.00 | -199.66629 | -199.67501 | 0.00000 | 0.00 |
| *other molecules* |  |  |  |  |  |  |  |  |  |  |  |
| $H_2CO$ | -114.56070 | -113.96557 | 0.59513 | -114.55669 | -114.56007 | -0.00063 | -0.11 | -114.54366 | -114.56051 | -0.00019 | -0.03 |
| $C_6H_6$ | -232.31284 | -230.13462 | 2.17822 | -232.33417 | -232.31137 | -0.00147 | -0.07 | -232.33896 | -232.32338 | 0.01054 | 0.48 |
| $NH_2CH_2COOH$ | -284.59469 | -283.04274 | 1.55195 | -284.59232 | -284.59624 | 0.00155 | 0.10 | -284.56033 | -284.60737 | 0.01268 | 0.82 |
| $NC_4H_4N$ | -264.42425 | -262.64628 | 1.77797 | -264.41541 | -264.40313 | -0.02112 | -1.19 | -264.41813 | -264.41845 | -0.00580 | -0.33 |
| $C_6H_5F$ | -331.65050 | -329.44169 | 2.20881 | -331.65332 | -331.64866 | -0.00184 | -0.08 | -331.64510 | -331.63188 | -0.01862 | -0.84 |
| $C_2F_2H_2$ | -277.25192 | -276.32568 | 0.92624 | -277.23997 | -277.26005 | 0.00813 | 0.88 | -277.25935 | -277.26288 | 0.01096 | 1.18 |

|  | E(exact) |  | atomic shift |  | atomic shift |
|---|---|---|---|---|---|
| H | -0.5 | H | 0.00430 | H | -0.00200 |
| C | -37.855770 | C | -0.00050 | C | 0.00460 |
| N | -54.611600 | N | -0.00146 | N | -0.00535 |
| O | -75.109800 | O | -0.01148 | O | -0.01744 |
| F | -99.807070 | F | -0.01384 | F | -0.00436 |

[a]Ref.2.

The results obtained for the few molecules tabulated in Tables 1 and 2 are encouraging since in all cases inclusion of the corr-exchange contribution brings the calculated total energy into much closer agreement with the exact value even when the initial CI error is quite large. We now consider a larger set of 40 molecules representing a range of bonding environments. Calculated energies from the two levels of CI treatment (full-basis and truncated virtual space) are depicted in Fig. 2 for calculations with and without corr-exchange corrections. Inclusion of the corr-exchange contribution is found to capture most of the missing correlation energy (> 98%) for each CI treatment bringing the calculated total energy into considerably improved agreement with the exact thermodynamic energy. The energy errors are further reduced on inclusion of the small atomic energy shifts defined in Table 2.

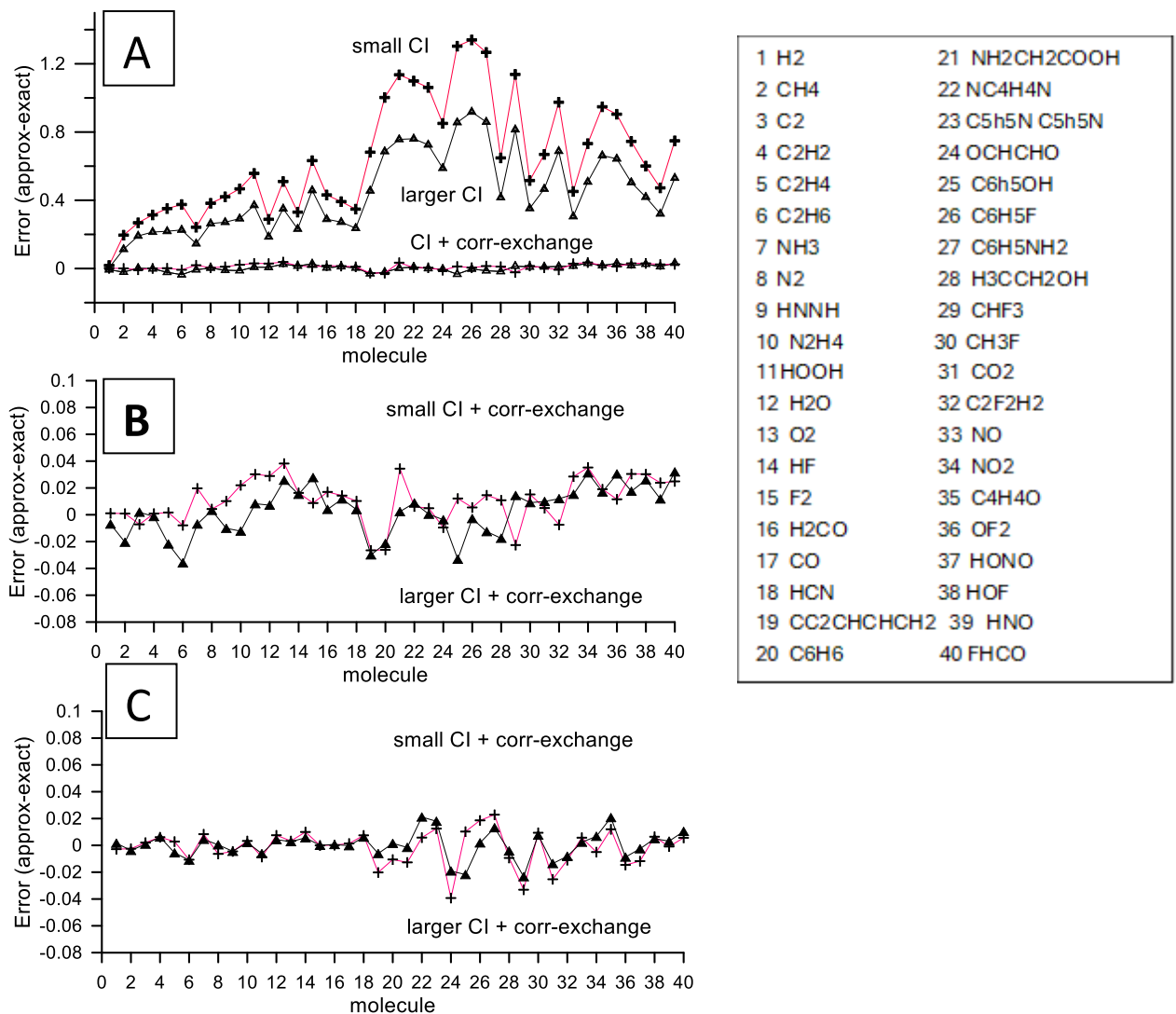

**Fig. 2.** Total energy errors from CI calculations on the molecules listed: 1) using the full virtual basis (larger CI) and 2) using a truncated virtual basis (small CI). A: Results are plotted for calculations with no corr-exchange contribution (small and larger CI) and for calculations including the corr-exchange contribution. B: An expanded scale is shown for the same calculations that include the corr-exchange contribution. C: Atomic constants are added to the corr-exchange energies. Energies are in hartrees. Parameters are the same as defined in Table 2.

Atomization energies provide a further measure of accuracy and the results depicted in Fig. 3, show errors remain small for almost all systems. As noted earlier, the error in atomization energy is the same as the error in total energy since the dissociation is to exact atoms. Larger percentage deviations occur when the atomization energy is small, for example, as in $OF_2$ (molecule 36 which has an atomization energy of 0.15 hartree. Summing over the absolute value of the errors in atomization energy gives average errors of 1.18% and 1.60% for the full-basis and truncated virtual space calculations, respectively,

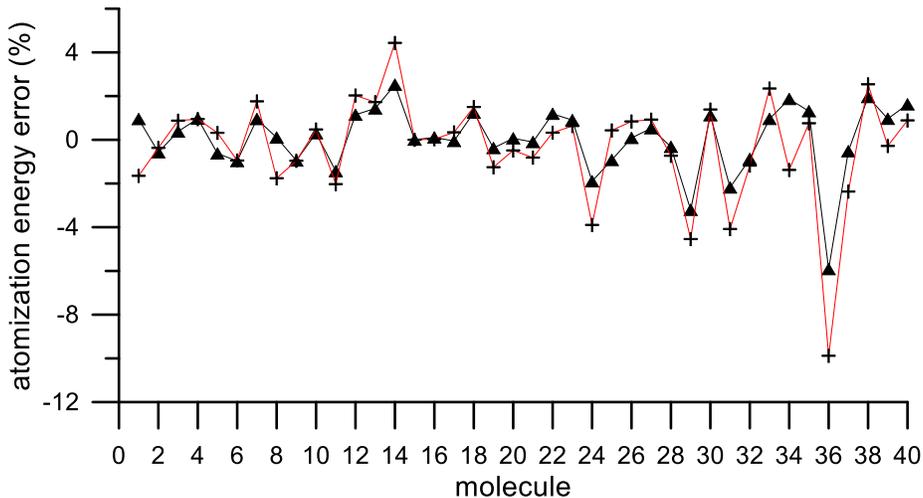

**Fig. 3.** Atomization energy errors from CI calculations on the molecules listed in Fig. 1: using the full virtual basis (larger CI) (black line) and using a truncated virtual basis (small CI) (red line). Total energy errors are in same as in Fig. 2 and are the same as atomization energy errors since the dissociation is to exact atoms. However, depicting the fraction or percent error is a sensitive measure that gives larger errors when the atomization energy is small as in molecule 36 (see text). The corr-exchange parameters and atomic constants are the same as reported in Table 2 $\gamma = 0.0176$ (full virtual space) $\gamma = 0.0254$ (truncated virtual space).

We now revisit the choice of all molecular orbital pairs regardless of spin to define the corr-exchange contribution. One might reasonably question this choice since excitations from opposite spin pairs provide larger contributions to the CI energy than excitations from same spin pairs. However, if we allow different weightings of the self-energy and contributions from different pairs the final energy lowering can be approximately the same. For example, consider a pair of doubly occupied spatial orbitals $\varphi_i$, $\varphi_j$. If we keep the self- energy as $-\gamma < \varphi_i(1) \varphi_i(2) | r_{12}^{-1} | \varphi_i(1) \varphi_i(2) >$ and change $\gamma$ by a factor of two for different-spin pairs and take $\gamma = 0$ for same-spin pairs the total energy lowering would be the same. We have investigated the limit of a zero contribution for same spin pairs for the same set of 40 molecules and find negligible improvements for S=0 molecules and a small decrease in accuracy for S=1 molecules.

It is interesting to compare the present method based on the exchange function with a previous approach that attempted to account for correlation and other errors by modifying the one-electron contribution, $<f_i|h|f_j>$ , over basis functions (JCP 2020)[5]. In that work, a constant $\gamma_M$ representing the energy defect is assigned to each type of atom $M$ and a one electron operator $h_i''$ that carries the defect contribution is added to the exact one-electron Hamiltonian $h$. The $h_i''$ is defined by its matrix elements

$$<f_i|h''|f_j> = \tfrac{1}{2}<f_i|f_j>(\gamma_M+\gamma_N)$$
$$f_i \; \varepsilon \; M \quad f_j \; \varepsilon \; N$$

Although absolute errors are slightly larger and there are more special cases in the earlier work, the overall comparison of atomization energy errors, depicted. in Fig. 4, shows comparable accuracy of the methods.

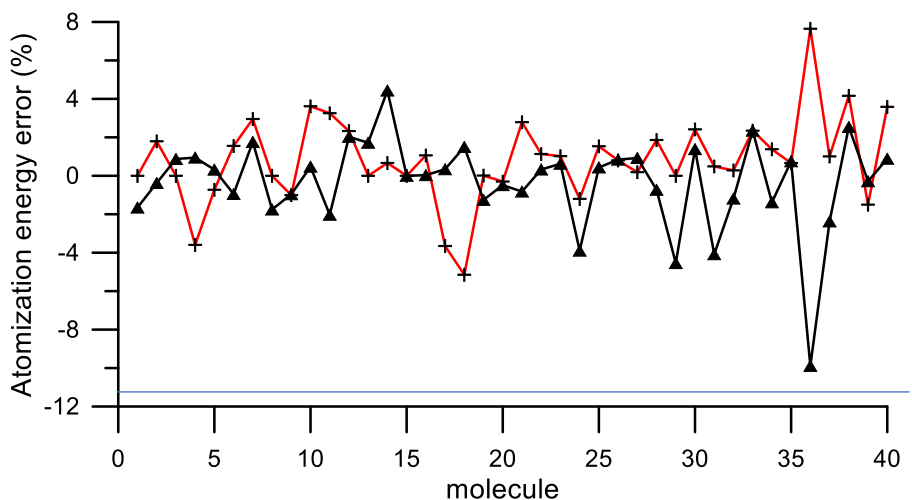

Fig. 4. Comparison of corr-exchange results (black line) with a previously reported method based on orbital overlap (red line). Both calculations are for the truncated virtual space. The small atomization error of OF$_2$ (molecule 36) magnifies the percent error.

We conclude by suggesting that a corr-exchange calculation could be useful to predict higher level CI energies from lower-level calculations and to provide a check on consistency. For example, it would be interesting determine how accurately coupled cluster results for a reference set of small molecules plus the corr-exchange contribution would predict CCSD(T) or higher-level CC energies for large molecules.

## V. Conclusions

A simple method for accounting for errors in CI calculations associated with missing excitations is proposed. The correlation energy and other errors associated with excitations from an electron pair $i, j$ are mapped onto an exchange interaction $-\gamma(\chi_i(1)\chi_j(2)|r_{12}^{-1}|\chi_j(1)\chi_i(2))$, $\gamma \geq 0$. involving all molecular orbital pairs regardless of spin. The parameter, $\gamma$, is determined from the exact thermodynamic energy of a few reference molecules and the calculated CI energy. The value of $\gamma$ depends on the basis and level of configuration interaction but is the same for all molecules. Calculated energies are compared with experimental thermodynamic data for a set of forty mainly organic molecules representing a wide range of bonding environments. Results are reported for two types of multi-reference CI calculations: 1) a triple-zeta basis plus d-type functions for C, N, O and F and an s, p basis for H, and 2) a severely truncated virtual space in which higher spherical harmonic basis functions are removed. Both CI treatments have large errors compared to exact energies. For the larger CI, including the corr-exchange contribution for $\gamma = 0.0176$ brings calculated CI energies for all systems into greatly improved agreement with experiment. The truncated virtual space produces many fewer configurations; however, increasing $\gamma$ to 0.0254 gives quite good energies. Including the corr-exchange contribution in a CI calculation can be accomplished simply by adding the contribution to the Coulomb interaction for pairs of spatial molecular orbitals

$$(\chi_i(1)\chi_j(2)|r_{12}^{-1}|\chi_i(1)\chi_j(2)>) - \gamma(\chi_i(1)\chi_j(2)|r_{12}^{-1}|\chi_j(1)\chi_i(2))$$

**Appendix**

**Basis set**

The basis for each atom is a near Hartree-Fock set of atomic orbitals plus extra two-component s- and p-type functions consisting of the two smaller exponent components of the atomic orbital; sets of two-component d and two-component p functions are added for first-row atoms and hydrogen, respectively. The latter d- and p-type functions were optimized by CI calculations on atoms. Orbitals are expanded as linear combinations of Gaussian functions: 1s(10), 2s(5), 2p(6), s′(2), s″(1), p′(2), p″(1), d(2), for C,N,O , 2p(7) for F, and 1s(4), s(1), p(2) for H where the number of Gaussian functions in each orbital is indicated in parentheses. No core potentials are used in the present calculations.

## Configuration interaction

All calculations are carried out for the full electrostatic Hamiltonian of the system

$$H = \sum_i^N [-\tfrac{1}{2}\nabla_i^2 + \sum_k^Q -\frac{Z_k}{r_{ik}}] + \sum_{i<j}^N r_{ij}^{-1}$$

A single-determinant self-consistent-field (SCF) solution is obtained initially for each state of interest. Configuration interaction wavefunctions are constructed by multi-reference expansions,[7-8]

$$\Psi = \sum_k c_k (N!)^{-1/2} det(\chi_1^k \chi_2^k \ldots \chi_N^k) = \sum_k c_k \Phi_k$$

In all applications, the entire set of SCF orbitals is used to define the CI active space. Virtual orbitals are determined by a positive ion transformation to improve convergence. Single and double excitations from the single determinant SCF wavefunction, $\Phi_r$, creates a small CI expansion, $\Psi'_r$,

$$\Psi'_r = \Phi_r + \sum_{ijkl} \lambda_{ijkl} \Gamma_{ij \to kl} \Phi_r = \sum_m c_m \Phi_m$$

The configurations $\Phi_m$, are retained if the interaction with $\Phi_r$ satisfies a relatively large second order energy condition

$$\frac{|<\Phi_m | H | \Phi_r>|^2}{E_m - E_r + \lambda} \geq 10^{-4}$$

The description is then refined by generating a large CI expansion, $\Psi_r$ by single and double excitations from all important members of $\Psi'_r$ to obtain

$$\Psi_r = \Psi'_r + \sum_m \left[ \sum_{ik} \lambda_{ikm} \Gamma_{i \to k} \Phi_m + \sum_{ijkl} \lambda_{iklm} \Gamma_{ij \to kl} \Phi_m \right]$$

where $\Phi_m$ is a member of $\Psi'_r$ with coefficient > 0.01. Typically, $\Psi'_r$ contains 200-400 dets. We refer to this expansion as a multi-reference CI. The additional configurations are generated by identifying and retaining all configurations, $\Phi_m$, that interact with $\Psi'_r$ such that

$$\frac{|<\Phi_m | H | \Psi'_r>|^2}{E_m - E_r + \lambda} \geq 10^{-6}$$

For the molecules investigated, approximately $10^5$-$10^6$ determinants occur in the final CI expansion, and the expansion can contain single through quadruple excitations from an initial representation of the state $\Phi_r$. The contribution of determinants not explicitly included along with size consistency corrections are estimated by perturbation theory. The value of $\lambda$ is determined so that the second order perturbation energy matches the CI value if first order coefficients

$$c_m = \frac{-<\Phi_m|H|\Psi'_r>}{E_m - E_r + \lambda}$$

are used for determinants in the CI calculation.

**Experimental thermodynamic data from NIST compilation** [a]

|  | Atomizaton energy vibrational kJ | ZPE cm-1 | integrated Cp kJ | Exact energy hartree | Exact atomization incl ZPE hartree |
|---|---|---|---|---|---|
|  |  |  | 6.197 |  |  |
| C |  |  | 6.536 | -37.85577 |  |
| N |  |  | 6.197 | -54.61160 |  |
| O |  |  | 6.725 | -75.10980 |  |
| F |  |  | 6.518 | -99.80707 |  |
| H |  |  | 6.197 | -0.50000 |  |
| H$_2$ | 432.1 | 2179.3 | 8.468 | -1.17447 | 0.17447 |
| C$_2$ | 600 | 914 | 10.169 | -75.94423 | 0.23269 |
| CH$_4$ | 1642 | 9480 | 10.016 | -40.52437 | 0.66860 |
| N$_2$ | 941.6 | 1165 | 8.670 | -109.58714 | 0.36394 |
| NH$_3$ | 1157.9 | 7214.5 | 10.043 | -56.58549 | 0.47389 |
| O$_2$ | 493.7 | 778 | 8.680 | -150.41118 | 0.19159 |
| H$_2$O | 917.8 | 4504 | 9.905 | -76.47989 | 0.37009 |
| F$_2$ | 154.5 | 447 | 8.825 | -199.67501 | 0.06088 |
| HF | 566.6 | 1980.7 | 8.599 | -100.53190 | 0.22483 |
| C$_2$H$_4$ | 2225.5 | 10784.7 | 10.518 | -78.60833 | 0.89679 |
| C$_2$H$_2$ | 1626.5 | 5660.5 | 10.009 | -77.35683 | 0.64529 |
| C$_2$H$_6$ | 2787 | 15853.5 | 11.884 | -79.84529 | 1.13375 |
| CO | 1071.80 | 1071.6 | 8.671 | -113.37868 | 0.41311 |
| H$_2$CO | 1495 | 5643.5 | 10.020 | -114.56070 | 0.59513 |
| HCN | 1265.7 | 3412.5 | 9.235 | -93.46499 | 0.49763 |
| NO | 627.1 | 938 | 9.192 | -129.96452 | 0.24312 |
| C$_6$H$_6$ | 5463 | 21392.5 | 14.331 | -232.31284 | 2.17822 |
| C$_4$H$_4$N$_2$ | 4488 | 16307.5 | 15 [e] | -264.42425 | 1.77798 |
| C$_5$H$_5$N | 5005.7 | 18909.2 | 15 [e] | -248.37746 | 1.98701 |
| NH$_2$CH$_2$COOH | 3885 | 17107 [b] | 15 [e] | -284.59469 | 1.55195 |
| C$_6$H$_5$NH$_2$ | 6211.7 | 24527 [c] | 16.184 [e] | -287.71772 | 2.47150 |
| FHCO | 1639.8 | 4449.6 | 10.02 [e] | -213.91366 | 0.64102 |
| CF$_2$CH$_2$ | 2338.5 | 7805 | 12.048 | -277.25192 | 0.92625 |
| C$_6$H$_5$F | 5580.2 | 19663.5 | 16.184 | -331.65050 | 2.20881 |
| CHOCHO | 2554.5 | 7868 | 13.673 | -227.93995 | 1.00881 |
| CH$_2$CHCHCH$_2$ | 4016.5 | 17998.5 | 15.134 | -156.03489 | 1.61181 |
| CH$_3$CH$_2$OH | 3182.50 | 16968.4 | 14.126 | -155.11080 | 1.28946 |
| C$_4$H$_4$O | 3977.4 | 14918.5 | 12.347 | -230.11576 | 1.58288 |
| C$_6$H$_5$OH | 5953.7 | 21798 [d] | 16.184 [e] | -307.60522 | 2.36080 |
| HOOH | 1055.5 | 5561.5 | 11.158 | -151.64696 | 0.42736 |
| HNNH | 1154.7 | 5947.7 | 9.997 | -110.69009 | 0.46690 |
| N$_2$H$_4$ | 1696.4 | 11204.9 | 11.449 | -111.92037 | 0.69718 |
| HNO | 823.7 | 2875 | 9.942 | -130.54823 | 0.32683 |
| HONO | 1253.3 | 4241.2 | 11.597 | -205.82788 | 0.49668 |
| CO$_2$ | 1598.00 | 2508 | 9.365 | -188.69544 | 0.62007 |
| CF$_2$ | 1050.3 | 1503.3 | 10.353 | -237.87679 | 0.40689 |
| OF$_2$ | 374.60 | 1110 | 10.895 | -274.87167 | 0.14774 |
| CH$_3$F | 1683.5 | 8376 | 10.135 | -139.84221 | 0.67937 |
| HOF | 634.2 | 2916.8 | 10.088 | -175.67171 | 0.25484 |
| CHF$_3$ | 1848.8 | 5457.5 | 11.565 | -338.50601 | 0.72904 |
| NO$_2$ | 927.70 | 1843 | 10.186 | -205.19294 | 0.36174 |

[a] Ref. 2   [b] CCpVDZ scaled   [c] SDCI 6-31G* scaled   [d] CC 6-31G* scaled

[e] Estimated from molecules with similar structure.

## Data Availability

Data used in this work vare available on request.

## References

1. Shavitt, Isaiah; Bartlett, Rodney J. (2009). Many-Body Methods in Chemistry and Physics: MBPT and Coupled-Cluster Theory. Cambridge University Press. ISBN 978-0-521-81832-2

**2.** NIST Computational Chemistry Comparison and Benchmark Database**,** NIST Standard Reference Database Number 101**,** Release 21, August 2020, Editor: Russell D. Johnson III http://cccbdb.nist.gov/ DOI:10.18434/T47C7

Kramida, A., Ralchenko, Yu., Reader, J., and NIST ASD Team (2021). *NIST Atomic Spectra Database* (ver. 5.9), [Online]. https://physics.nist.gov/asd [2022, May 5]. National Institute of Standards and Technology, Gaithersburg, MD. DOI: https://doi.org/10.18434/T4W30F